\documentclass[10t]{article}
\usepackage{amsmath}
\usepackage[]{graphicx}
\usepackage[T1]{fontenc}
\usepackage{amssymb}
\usepackage{amscd}
\title{Double exchange model for correlated electrons in systems with $t_{2g}$ orbital degeneracy}
\author{Krzysztof Wohlfeld \\ \small {Marian Smoluchowski Institute of Physics, Jagellonian University,} 
\\ \small {Reymonta 4, PL-30059 Krak\'ow, Poland} }
\begin{document}
\date{}
\maketitle
\begin{abstract}
We formulate the double exchange (DE) model for systems with $t_{2g}$ orbital degeneracy, relevant for 
hole-doped cubic vanadates. In the relevant regime of strong on-site Coulomb 
repulsion $U$ we solve the model using two distinct mean-field approximations: Hartree-Fock,
and mean-field applied to the formulation of the model in the Kotliar-Ruckenstein slave boson represenation. 
We show how, due to the relative weakness of the DE mechanism via $t_{2g}$ degenerate orbitals, the anisotropic $C$-type
antiferromagnetic metallic phase and the orbital liquid state can be stabilized. This is contrasted with the DE mechanism 
via $e_g$ degenerate orbitals which stabilizes the ferromagnetic order in the hole-doped regime of manganites. 
The obtained results are in striking agreement with the observed 
magnetic structures and the collapse of the orbital order in the doped La$_{1-x}$Sr$_x$VO$_3$, Pr$_{1-x}$Ca$_x$VO$_3$ and Nd$_{1-x}$Sr$_x$VO$_3$.    
\end{abstract}
\section{Introduction}
Recently orbital degrees of freedom in the strongly correlated electron systems
have attracted much attention \cite{Tok00} due to their crucial role in the stability
of the various magnetic phases in the Mott insulators \cite{Fei99} or in the partial 
explanation of the Mott metal-insulator transition itself \cite{Dag04}. Typical examples 
of such systems are the transition-metal oxides with partly filled $e_g$ or $t_{2g}$ 
degenerate orbitals. The remarkable feature of these oxides is the large on-site Coulomb repulsion $U\gg t$, 
with $t$ being the effective electron hopping. 
In case of the stoichiometric (undoped) oxides the Coulomb repulsion between electrons 
supresses charge 
fluctuations and leads to the effective low-energy $\propto\! J\!\!=\!\!4t^2\!/U$ 
superexchange (SE) interactions between spin {\itshape and} orbital degrees of freedom. 
In addition, the atomic Hund's interaction $J_H\ll U$ aligns the spins of the electrons occupying the  
degenerate and almost degenerate orbitals on the same site and should be taken into account in the realistic SE-type models. 
Interestingly, it is the Hund's interaction 
which to large extent stabilizes
particular spin (magnetic) {\itshape and} orbital order. It should be stressed that though the $e_g$
orbitals usually order as in e.g. LaMnO$_3$ \cite{Fei99}, it is rather not the case of the $t_{2g}$ orbitals which can form a disordered 
quantum orbital liquid (OL) as in e.g. LaVO$_3$ \cite{Kha01}. The different kinds of the orbital state are concomitant with various anisotropic magnetic phases 
such as $A$-AF [ferromagnetic (FM) planes staggered antiferromagnetically (AF) in the other direction] in LaMnO$_3$ \cite{Fei99} 
or $C$-AF [FM chains staggered AF in the other
two directions] in LaVO$_3$ \cite{Kha01}.  

Doping with holes destroys the insulating state, modifies the SE interaction and hence can also modify the magnetic order and orbital
state stabilized in the undoped case. In such systems the motion of holes is strongly affected by: (i) intersite SE interaction $\propto J$, 
(ii) on-site Hund's interaction $\propto J_H$, which is captured by the Kondo-lattice model. In the realistic limit in the transition metal oxides
$J_H\gg t\gg J$ this model reduces to the double exchange model (DE) \cite{Zen51}. However, for the above discussed class of the transition-metal oxides, 
orbital degeneracy needs also to be taken into account in 
such a model. This was done in the seminal work of van den Brink and Khomskii \cite{Bri99} for electrons with the $e_g$ degeneracy, leading to a naively counterintuitive 
picture of the DE mechanism. One can thus pose a
question how the DE mechanism would be modified in the case of the $t_{2g}$ orbital degeneracy, as the SE interactions for the undoped oxides differ qualitatively 
for the $t_{2g}$ and $e_g$ cases. Hence, in this paper we want to answer three questions: (i) what is the nature of the magnetic order stabilized by the DE mechanism
for correlated electrons with $t_{2g}$ orbital degeneracy, (ii) what is the nature of the orbital order, and (iii) how do these results differ from the $e_g$ and the 
nondegenerate case. 

The paper is organized as follows. In the following chapter we introduce the DE Hamiltonian with $t_{2g}$ orbital degrees of freedom. Then we solve the model 
using two different mean-field approximations and we show that the metallic state coexists with the $C$-AF order and OL state for the broad range of hole-doping. 
Next we discuss the results, i.e.: the validity of the approximations, the generic role of the $t_{2g}$ orbitals in the stability of the above phases, the physical relevance 
of the model by comparison with the experiment. The paper is concluded by stressing: the distinct features of the DE mechanism via $e_g$ and via $t_{2g}$
degenerate orbitals, and the crucial role of the Coulomb repulsion $U$.
\section{Realistic double exchange Hamiltonian}
\label{sec:2}
We start with the realistic semiclassical DE Hamiltonian with $t_{2g}$ orbital degrees of freedom, relevant for the hole-doped cubic vanadates \cite{Woh06}:
\begin{equation}
\label{eq:2}
{\cal H}_{\rm DE} = 
-  u_y \sum_{ i, j \|  \hat{y}} \widetilde{a}^\dagger_i \widetilde{a}_j \ 
-  \sum_{ i, j  \| \hat{z}} \widetilde{a}_i^\dagger \widetilde{a}_j  \ 
-  u_x \sum_{ i, j \|  \hat{x}} \widetilde{b}^\dagger_i \widetilde{b}_j \ 
-  \sum_{ i, j  \| \hat{z}} \widetilde{b}_i^\dagger \widetilde{b}_j \ 
+J\!\!\sum_{\langle ij \rangle  \|  \hat{x} , \hat{y}}\! 
          \text{\bf{S}}_{i}\cdot \text{\bf{S}}_{j}.
\end{equation}
where: the hopping amplitude $t\!\!=\!\!1$; the restricted fermion creation operators
$ {\widetilde{a}}^\dagger_i = a^\dagger_i(1-b_i^\dagger b_i)$, 
$ {\widetilde{b}}^\dagger_i = b^\dagger_i(1-a_i^\dagger a_i)$, where 
$a^\dagger_i(b^\dagger_i)$ creates a spinless electron at site $i$ in 
$yz(zx)$ orbital; and  $\text{\bf{S}}_{i}$ are core spin $S=1/2$ operators of $t_{2g}$ 
electrons in occupied $xy$ orbitals. We introduce the $u_x$ and $u_y$ variational parameters 
to be defined later. The Hamiltonian has 
the following features: (i) the first four terms describe the kinetic energy of 
the electrons in the degenerate $yz$ and $zx$ orbitals which can hop only in the allowed $(\hat y, \hat z)$ or $(\hat x, \hat z)$ plane to 
the nearest neighbour (nn) site $i$ providing there are no other 
electrons at site $i$ in these orbitals ($U=\infty $ assumed implicitly), (ii) the last term describes the AF coupling 
between core spins at nn sites in the $(\hat x, \hat y)$ plane due to the SE interaction originating from 
the electrons in the always occupied $xy$ orbitals ($n_{xy}=1$ during doping), (iii) the 
SE interactions due to the itinerant electrons in $yz$ and $zx$ orbitals are neglected.

The variational parameters are defined as $u_x=\cos(\theta_x/2)$ and $u_y=\cos(\theta_y/2)$, where $\theta_x$
$(\theta_y)$ is the relative angle between core spins in the $\hat{x}$ $(\hat {y})$ direction. This follows from the Hund's rule which aligns the spins of the 
itinerant electrons with the core spins and which is not explicitly written in Eq. (\ref{eq:2}) but enters via $u_x$ and $u_y$ parameters. Then the last term of 
Eq. (\ref{eq:2}) can be written as:
\begin{equation}
\label{eq:3}
{E}_{\rm SE} = J\!\sum_{\langle ij \rangle  \|  \hat{x} , \hat{y}}\! 
          \text{\bf{S}}_{i}\cdot \text{\bf{S}}_{j}=\frac{J}{2}L^3(u_x^2\!+\!u_y^2\!-\!1), 
\end{equation}
where $L^3$ is the number of sites in the crystal. Eq.(\ref{eq:2}) together with Eq. (\ref{eq:3}) shows the competition 
between the DE mechanism allowing for the hopping in the underlying
FM background with the SE interaction supporting the AF order. 
\section{Numerical results}
The ground state of Eq. (\ref{eq:2}) was found using two distinct mean-field approximations: 
Hartree-Fock, and mean-field applied to the formulation of the model Eq. (\ref{eq:2}) in the Kotliar-Ruckenstein slave boson representation \cite{Kot86}.
\subsection{Hartree-Fock approximation (HFA)}
To facilitate the treatment of the hole doped electronic systems we enlarge the original  Fock space, which is relevant for the Hamiltonian Eq. (\ref{eq:2}),
by adding a boson creation (annhilation) operator $e^\dag_i(e_i)$, which creates (annihilates) a charged hole. Then, in order to get rid of the nonphysical
states in the enlarged Fock space we need a {\itshape local} constraint:
\begin{equation}
\forall_i \quad a^\dag _i a_i+b_i^\dag b_i+e_i^\dag e_i=1.
\label{constrainthf}
\end{equation}
Note that the boson controlling double occupancies is not introduced since such configurations are forbidden
for $U=\infty $. 
In order  to solve Eq. (\ref{eq:2}) with the constraint Eq. (\ref{constrainthf}) in a simple way, we introduce mean-field approximation 
for the above constraint and replace the following combination of fermion operators by averages: 
\begin{align}
&\forall_{i}\quad  b^\dagger_i b_i \rightarrow \langle b^\dagger_i b_i \rangle=: \delta \ \text{and} \ 0\leq \delta\leq 1,   \label{MFa} \\
&\forall_{i}\quad  a^\dagger_i a_i \rightarrow \langle a^\dagger_i a_i \rangle=1-x-\delta \ \text{and} \ 0\leq \delta+x\leq 1, 
\label{MFb}
\end{align}
where in Eq. (\ref {MFb}) we use constraint Eq. (\ref{constrainthf}), where $\langle e^\dagger_i e_i \rangle =x$ 
and $x$ is the number of doped holes per site. Besides, the inequalities in Eq. (\ref {MFa}) and 
(\ref {MFb}) follow from the fact that we are discussing the hole-doping regime, i.e. with not more than one itinerant electron per site. We call
this mean-field approximation an HFA approximation since it is equivalent to
decoupling the interaction term in an HFA way:
\begin{equation}
(U\!-\!3J_H)\  a ^\dag _i a_i b^\dag_i b_i \simeq (U\!-\!3J_H) \left( \langle a^\dag_i a _i \rangle b^\dag _ib_i \!+\! a^\dag_i a _i \langle b^\dag _i b_i \rangle\! -\! \langle a_ia^\dag _i\rangle \langle b^\dag _ib_i \rangle \right),
\label{HF}
\end{equation}
and then making the $U\rightarrow \infty $ limit [where $U-3J_H$ is the energy of the $t_{2g}^2$ high spin state]. 
Note that the above used HFA does not allow for the alternating orbital (AO) order but we
do not expect such an order to persist during doping. This is because the kinetic energy of the electrons in the AO is rigorously zero 
for $U=\infty $ {\itshape and} the vanishing hopping between different orbitals \cite{Fei05}. 
\begin{figure}[t]
\includegraphics[width=\textwidth]{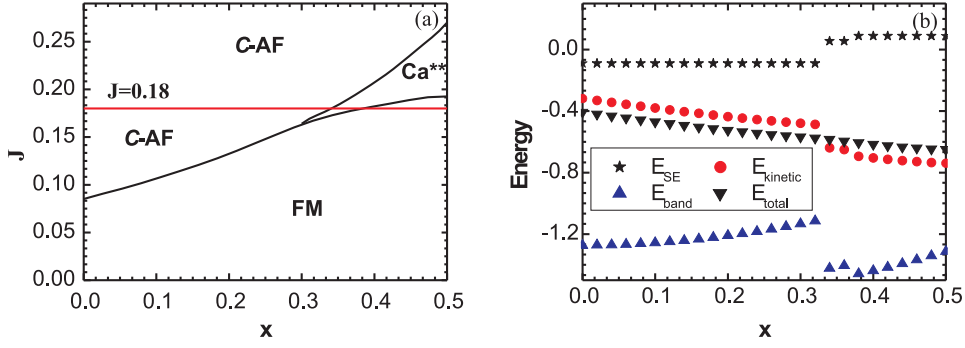}
\caption{\small{Results of the minimisation of Eq. (\ref{E_MF}) (with the constraint Eq. (\ref{Fermi})): 
(a) Magnetic phase diagram at $T=0$. Under increasing hole doping $x$, the $C$-AF order changes via the canted Ca** phase, 
with spins canted on two bonds, to a homogenous FM phase. The grey horizontal line depicts the realistic value of SE $J= 0.18$ in LaVO$_3$; 
(b) Competing: correlated kinetic energy
($E_{\rm kinetic}$), and magnetic energy ($E_{\rm SE}$) as a function of hole doping $x$ for the fixed value of SE $J=0.18$. Kinetic energy for 
uncorrelated electrons ($E_{\rm band}$) and the total energy of the system ($E_{\rm total}$) is also shown.}}
\label{fig:propermean}
\end{figure} 
Then $\mathcal{H}_{\rm DE}$ can be diagonalized and the total energy of the system in the mean-field approximation is:
\begin{align}
E_{\rm HFA}/L\!= \! (\!1\!\!-\!\!\delta)^2\! \sum_{ \substack{(k_y, k_z) \in \\ S_F(u_y, 1\!-\!x\!-\!\delta)}}\!\!\! \mathcal{E}(k_y, k_z; u_y) 
\!+\! (\!x\!\!+\!\!\delta)^2\! \sum_{ \substack{(k_x, k_z) \in \\ S_F(u_x, \delta)}} \mathcal{E}(k_x, k_z; u_x) 
+ E_{\rm SE}
\label{E_MF}
\end{align}
where the kinetic energy renormalizing factors follow from the HFA. 
The sums go over the mesh of $\bf{k}$'s belonging to the Fermi volume $S_F$ which "accommodates"
$1-x-\delta$ or $\delta$ electrons per site (respectively) and whose shape depends also on 
the band dispersion relations for the tight-binding model renormalized by the underlying magnetic order:
\begin{align}
\mathcal{E}(k_y, k_z; u_y)\!=\!-2(u_y\cos k_y\!+\!\cos k_z) \  {\rm and}  \ \mathcal{E}(k_x, k_z; u_x)\!=\!-2(u_x\cos k_x\!+\!\cos k_z).
\label{E}
\end{align}
Since we are calculating separately the energies of the $yz$ and $zx$ bands we need an additional constraint for the Fermi levels 
$\mathcal{E}_F$ as the two bands fill up gradually:
\begin{equation}
\mathcal{E}_F(1-x-\delta, u_y)=\mathcal{E}_F(\delta, u_x).
\label{Fermi}
\end{equation}

For fixed values of SE $J$ and hole doping $x$ we minimise numerically the total energy
Eq. (\ref{E_MF}) (with the constraint Eq. (\ref{Fermi})) with respect to the $u_x$ and $u_y$ variational parameters, which determine the magnetic structure. 
The resulting magnetic phase diagram at $T=0$ is shown in Fig. \ref{fig:propermean}(a).  In Fig. \ref{fig:propermean}(b) we show 
the competing kinetic and magnetic (SE) energies as a function of hole doping $x$ for the realistic value of SE $J=0.18$ and obtained in the
minimum of the total energy. Let us also note that for {\itshape all} of the discussed values of $J$ and $x$ it was found 
that $\delta^{min}=\langle c^\dagger_i c_i \rangle=\langle d^\dagger_i d_i \rangle=(1-x)/2$,  i.e. the number of electrons 
in both orbitals is the same and the ferrorbital order (FO) is not stabilized. 

\subsection{Kotliar-Rukenstein slave boson approach (KRA)}
We rewrite the Hamiltonian Eq. (\ref{eq:2}) using Kotliar-Ruckenstein slave boson representation \cite{Kot86} adopted to the orbital case. 
So we enlarge the Fock space by introducing three auxiliary boson fields and decouple the electron creation 
operator $a^{\dagger}_i(b^{\dagger}_i)$ into a fermion creation operator $f^{\dagger}_i(g^{\dagger}_i)$ carrying the orbital degree of freedom (orbital flavour), 
a boson creation operator $p^\dag_i(q^\dag_i)$, and a boson annihilation operator $e_i$ carrying the charge degree of freedom (i.e. $e^{\dagger}_i$ creates a charged hole):
\begin{equation}
a^{\dagger}_i=f^{\dagger}_iz^\dag _i, \quad
b^{\dagger}_i=g^{\dagger}_iy^\dag_i, 
\label{sb22}
\end{equation}
where
\begin{align}
z^\dag_i\!:=\!\frac{p^\dag_ie_i}{\sqrt{(1-e^\dag e_i\! -\! q_i^\dag q_i)(1\!-\!p_i^\dag p_i)}}, \ \ {\rm and} \ \ 
y^\dag_i\!:=\!\frac{q^\dag_ie_i}{\sqrt{(1-e^\dag e_i\! -\! p_i^\dag p_i)(1\!-\!q_i^\dag q_i)}}.
\label{z}
\end{align}
Applying this to the Hamiltonian $\mathcal{H}_{\rm DE}$ we get:
\begin{align}
\mathcal{H}_{\rm DE}= &- u_y \sum_{ i, j   \|  \hat{y}} {f}^\dag_{i } {f}_{j }z^\dag_i z_j
-  \sum_{ i, j  \|  \hat{z}} {f}^\dag_{i } {f}_{j }z^\dag_i z_j + E_{\rm SE} \nonumber \\
&-  u_x \sum_{ i, j   \|  \hat{x}} {g}^\dag_{i} {g}_{j} y_i^\dag y_j
-  \sum_{ i, j   \|  \hat{z}} {g}_{i}^\dag {g}_{j} y_i^\dag y_j , 
\label{H^SB2_UHA}
\end{align}  
where we also need constraints to get rid of the nonphysical states in the enlarged Fock space:
\begin{align}
\forall_i \quad p^\dag _i p_i+&q_i^\dag q_i+e_i^\dag e_i=1, \quad 
\forall_i \quad p^\dag _i p_i=f_i^\dag f_i \quad \text{and} \quad \forall_i \quad q_i^\dag q_i=g_i^\dag g_i.
\label{cons}
\end{align}
 
Next we introduce mean-field approximation:
\begin{align}
\forall_i \quad z_i^\dag z_i\rightarrow \langle z_i^\dag z_i\rangle=\frac{\langle p^\dag_i e_i e_i^\dag p_i \rangle}
{\langle 1-e^\dag e_i - q_i^\dag q_i \rangle \langle 1-p_i^\dag p_i \rangle }=\frac{x}{x_a}, \nonumber \\ 
\forall_i \quad y_i^\dag y_i\rightarrow \langle y_i^\dag y_i\rangle=\frac{\langle q^\dag_i e_i e_i^\dag q_i \rangle}
{\langle 1-e^\dag e_i - p_i^\dag p_i \rangle \langle 1-q_i^\dag q_i \rangle }=\frac{x}{x_b},
\label{constraintmf2}
\end{align}
where we use Eq. (\ref{cons}) and introduce: $\langle1- p_i^\dag p_i \rangle=x_a$ the mean number of holes per site in orbital $yz$, 
$\langle 1- q_i^\dag q_i \rangle=x_b$ the mean number of holes per site in orbital $zx$, and $\langle e^\dagger_i e_i \rangle =x$ the mean number of holes per site. 
Furthermore, introducing parameter $\delta$ defined in 
the same manner as in Eq. (\ref{MFa}-\ref{MFb}) yields $x_a=x+\delta$ and $x_b=1-\delta$. Then the total energy of the system in the mean-field approximation is:
\begin{align}
E_{\rm KRA}/L=  \ \frac{x}{x+\delta} \!\!\! \sum_{ \substack{(k_y, k_z) \in \\ S_F(u_y, 1-x-\delta)}} \!\!\!\mathcal{E}(k_y, k_z; u_y) 
+ \frac{x}{1-\delta} \!\!\! \sum_{ \substack{(k_x, k_z) \in \\ S_F(u_x, \delta)}}\! \mathcal{E}(k_x, k_z; u_x) + E_{\rm SE},  
\label{E^SB2_MF}
\end{align}
where the electronic bands are defined in the same manner as in Eq. (\ref{E}) and the constraint Eq. (\ref{Fermi}) also holds. 

\begin{figure}[t]
\includegraphics[width=\textwidth]{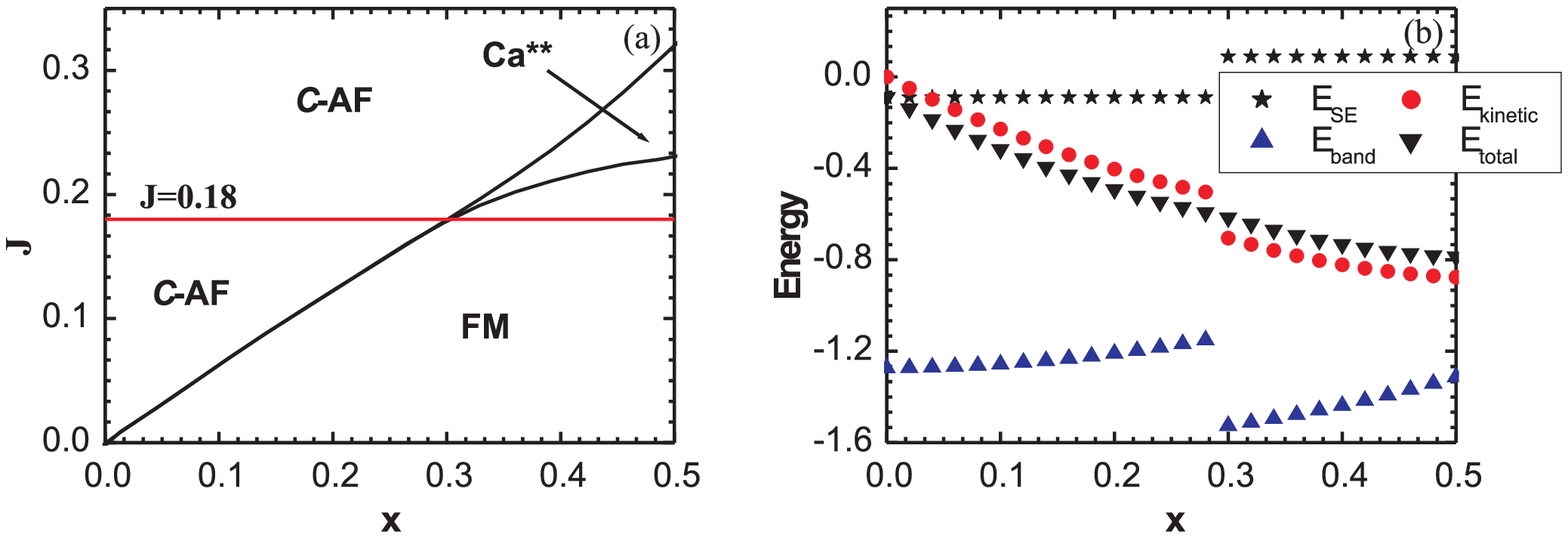}
\caption{\small{Results of the minimisation of Eq. (\ref{E^SB2_MF})  (with the constraint Eq. (\ref{Fermi})):
(a) Magnetic phase diagram at $T=0$. Under increasing hole doping $x$, 
the $C$-AF order changes via the canted Ca** phase, 
with spins canted on two bonds, to a homogenous FM phase. The grey horizontal line depicts the realistic value of SE $J= 0.18$ in LaVO$_3$.
Note that for $x=0$ the obtained magnetic order consists of decoupled AF $(\hat x,\hat y)$ planes [after Ref. \cite{Woh06}]; (b) Competing: correlated kinetic energy
($E_\text{kinetic}$), and magnetic energy ($E_\text{SE}$) as a function of hole doping $x$ for the fixed value of SE $J= 0.18$. Kinetic energy for 
uncorrelated electrons ($E_\text{band}$) and the total energy of the system ($E_\text{total}$) is also shown.}}
\label{fig:properslave2}
\end{figure}

As in the HFA we deduce the resulting magnetic structures by minimising the total energy Eq. (\ref{E^SB2_MF}) with the constraint Eq. (\ref{Fermi}).
The resulting magnetic phases are shown in Fig. \ref{fig:properslave2}(a) and the competing energies at fixed SE $J$ as a function of hole doping $x$     
are shown in Fig. \ref{fig:properslave2}(b). We should also stress that the number of electrons in both
orbitals does not change during doping and is equal to $\delta^{min}=(1-x)/2$. Hence, the kinetic energy renormalizing factors are equal to $2x/(x+1)$, 
the well-known Gutzwiller factors.
\section{Discussion of the results}
\subsection{Why Hartree-Fock approximation fails}
Let us first discuss the validity of the two mean-field approximations. For the HFA we infer that 
for small doping ($x \rightarrow 0$) the correlated kinetic energy ($E_\text{kinetic}$ in Fig. \ref{fig:propermean}(b)) 
does not tend to zero and instead equals 1/4 of the uncorrelated kinetic energy for $x=0$. 
This is a remnant of the fact that in the HFA the probability for existence of the hole at each site is equal to 25\%. 
Thus, we conclude that the configurations with double occupancies are not suppressed and that HFA fails, at least in the 
low hole doping regime. The reason why we cannot use HFA, unlike in e.g. the undoped antiferromagnet, is that the orbital state, which is favoured here by 
the DE mechanism, is not an ordered state. If we were to get the FO or AO state, HFA would give physically relevant results.
\subsection{Stability of the $C$-AF: generic role of the degenerate $t_{2g}$ orbitals}
We concentrate on the results obtained within KRA which give the (qualitatively) right value of the correlated kinetic energy of the Hamiltonian Eq. (\ref{eq:2}), 
cf. Fig. \ref{fig:properslave2}(b).  The $C$-AF order and a concomitant one-dimensional (1D) metallic state is stable
for a broad range of the phase diagram parameters, in particular for the realistic value of SE $J=0.18$ and $0<x< 0.28$ hole doping.
This means that the DE mechanism does not win in the $(\hat x, \hat y)$ planes, where the SE mechanism takes over. The first explanation of 
this phenomenon would be to ascribe the role of the driving force for the stability of the $C$-AF order to the Coulomb repulsion $U$: 
the correlated kinetic energy ($E_\text{kinetic}$ in Fig. \ref{fig:properslave2}(b)) is much smaller than the bare band energy 
($E_\text{band}$ in Fig. \ref{fig:propermean}(b)) and hence the possible kinetic energy gain in the homogeneous FM state is 
much smaller in the correlated regime than in the conventional free electron regime. 
However, for hole-doped manganites $U$, $J_H$ and $t$ are of the same order as for hole-doped vanadates \cite{Miz96}. 
But in the manganites for $x>  0.18$ hole doping the homogeneous FM order, concomitant with the metallic state, is stabilized mainly due to the DE mechanism \cite{Dag01}!
The only difference is that in the manganites the itinerant electrons hop between degenerate $e_g$ orbitals and in the vanadates between degenerate $t_{2g}$ orbitals.
Besides, the AF interaction between core spins is truely three-dimensional in manganites and not two-dimensional (2D) as it is the case of our model. Thus the relative weakness of the DE mechanism 
in the Hamiltonian Eq. (\ref{eq:2}) can 
be attributed to the specific features of the $t_{2g}$ electrons: the strictly 2D and strictly flavor conserving hopping between these degenerate orbitals \cite{Woh06}.

This generic role of the $t_{2g}$ orbitals in the stability of the $C$-AF order is threefold:

(i) The AF SE interactions are rigorously 2D because for $xy$ electrons the virtual hopping processes leading to the SE interactions can happen only in 
the $(\hat x, \hat y)$ plane. Hence it makes possible: the hopping in the $\hat z$ direction without destruction of the magnetic order, 
and thus the gain of the kinetic energy without 
any loss in the magnetic energy. Though this is rather a specific feature of our simplified SE interactions than of the DE mechanism itself it should be stressed that 
the physical setting in which we can have the DE via $t_{2g}$ degenerate orbitals is such that the AF core spin interactions are 2D.

(ii)  The degeneracy of the orbitals reduces the correlated kinetic energy and hence makes it energetically unfavourable to get homogenous FM order. 
Cf. Fig. 2(a) of Ref. \cite{Woh06} where for the system without orbital degeneracy with holes doped {\itshape only} into $zx$ orbital, i.e. with the Hamiltonian:
\begin{equation} 
\mathcal{H}_{zx}= -  u_x \sum_{ i, j   \|  \hat{x}} {d}_{i}^\dag {d}_{j} 
-  \sum_{ i, j   \|  \hat{z}} {b}_{i}^\dag {b}_{j} + \frac{J}{2}L^3(u_x^2+u_y^2-1), 
\label{zx}
\end{equation}
we obtain that for the realistic value of $J=0.18$ the FM order is already stable for hole doping $x>0.1$. 
Note that this comparison is valid only if the degenerate orbitals are in 
the OL state which indeed is the case, cf. discussion below.

(iii) Since each of the $t_{2g}$ orbitals has a {\itshape different} inactive axis [i.e. the cubic direction in which the electrons cannot hop] this means that if one of these
orbitals is "engaged" in the AF SE interactions [cf. (i)] than the other two orbitals have one {\itshape active} axis in the direction perpendicular to the plane with AF bonds. 
Hence, hopping in the $\hat z$ direction, assumed in point (i), is indeed possible.   
\subsection{Stability of the orbital liquid}
The orbital state which is concomitant with the obtained magnetic phases is the OL since: (i) we obtained $x_a=x_b$ for all of the magnetic 
phases, (ii) $x_a=x_b \Rightarrow 
\langle T_i^z\rangle\! :=\! \langle p^\dag_i p_i \rangle \!-\! \langle q^\dag_i q_i \rangle\! = \!0$ where $T_i^z$ is the $z$ component of the pseudospin 
orbital operator \cite{Fei05},
(iii) $\langle a^\dag_i b_i \rangle\! =\! \langle b^\dag_i a_i \rangle \!=\!0 \Rightarrow \langle T_i^x \rangle \!= \! \langle T_i^y\rangle \! =\!0$ for the $x$ and $y$ components
of the pseudospin orbital operator \cite{Fei05}. The reason why none of the ordered state are realized in nature is twofold. Firstly, as already described, the AO state
prohibits hopping due to the vanishing hopping integrals between different orbitals. This mimics the confinement of holes in the Neel AF, 
but stays in contrast with the behaviour of the $e_g$ orbitals. 
Thus we cannot gain any kinetic energy upon doping and this state is unphysical. Secondly, the FO state is also unphysical. A priori, due to the Pauli Principle the electrons
in the polarized state can better avoid each other than in the unpolarized one. Hence the correlated kinetic energy of the FO state is enhanced in comparison with the OL, cf. Fig. 8 
(with $\gamma=0$) of Ref. \cite{Fei05}. Though, the condition of the equilibrium for the whole system yields that 
the chemical potentials (Fermi energies) of the two fermionic subsystems 
(one with electrons in $yz$ orbitals and the other one in $zx$ orbitals) should be equal. 
For the FO state to be realized in nature it means that the DE mechanism should raise the bottom 
of one band so that the electrons can pour into the other one. However, the DE is too weak in our system and this does not happen. 
\subsection{Physical relevance of the model: comparison with the experiment}
As already pointed out our model describes the physics present in the hole-doped cubic vanadates such as La$_{1-x}$Sr$_x$VO$_3$ 
where the following phases were observed experimentally upon changing the hole doping $x$ in $T\rightarrow 0$: 
$C$-AF, AO and insulating one for $x<0.178$ hole concentration, 
$C$-AF and metallic one for $0.178<x<0.26$, and paramagnetic and 
metallic one for $0.26<x<0.327$ \cite{Fuj05}. 
Rather similar phases were observed experimentally in Pr$_{1-x}$Ca$_x$VO$_3$ and Nd$_{1-x}$Sr$_x$VO$_3$ \cite{Fuj05}. 
Hence our results stay in well agreement with the experiment, as we did not expect to explain by our model Eq. (\ref{eq:2}) 
the filling control metal-insulator transition present in the system, but merely the coexistence of  the metallic phase, the OL state 
and the $C$-AF magnetic order in the above mentioned
doping range. However, let us note that including the SE interactions between the itinerant electrons would stabilize the 1D OL in the $\hat z$ direction and AO in 
the $(\hat x, \hat y)$ plane \cite{Kha01}. Thus, if
we included the realistic SE in the Hamiltonian Eq. (\ref{eq:2}) we would get the insulating behaviour for finite hole doping due to the AO order in the plane which would act 
as a string potential for holes moving in the $\hat z$ direction.
This suggests that further investigation of the {\itshape even more realistic} DE model with $t_{2g}$ orbital degrees of freedom
{\itshape and} SE interactions between itinerant electrons would not only uncover some new interesting physics but also could explain the experimentally observed phases
in the doped cubic vanadates.
\section{Summary}
In summary, we have shown the distinct features of the {\itshape double exchange via $t_{2g}$ degenerate} orbitals. 
These features, attributed to the $t_{2g}$ orbital degeneracy, are the following: 

(i) Similar to the electron-doped $e_g$ degenerate orbitals \cite{Bri99} the orbital degeneracy {\itshape naturally allows} for the coexistence of the anisotropic 
magnetic phases and the metallic phase. This is a generic feature of the DE mechanism via degenerate orbitals. 

(ii) In contrast with the hole-doped $e_g$ case \cite{Dag01} the homogenous FM order is not stable for large range of doping in the discussed here hole-doped regime 
of $t_{2g}$ case. This is due to the different geometry of the $t_{2g}$ orbitals causing the hopping being strictly 2D and strictly flavour conserving, hence weakening 
the DE mechanism. 

(iii) Upon hole-doping the orbitals form the OL state due to the weakness of the DE mechanism. Hence, they do not order as it is typical for the electron-doped 
$e_g$ systems \cite{Bri99}. However, it should be stressed that the type of the orbital state for hole-doped $e_g$ systems is still under discussion \cite{Fei05}. 

Besides, the {\itshape correlated} behaviour of the itinerant electrons plays {\itshape a crucial role} in the stability of the obtained magnetic and orbital phases.
This is especially dramatic in the orbital sector, where the orbitals form the orbital liquid and the HFA [suitable either for weakly interacting electrons or for 
the ordered states] fails qualitatively in giving the right values of the kinetic energy. 
Altogether these are striking results which can serve as a basis for further investigation of the correlated phenomena in hole-doped systems with $t_{2g}$ 
orbital degeneracy.

{\bf Acknowledgments} I am particularly grateful to A. M. Ole\'s for his invaluable idea and discussions. 
I would like to acknowledge support by the Polish Ministry of Science and Education under Project No.~1 P03B 068 26.


\begin{thebibliography}{10}
\bibitem{Tok00} Y. Tokura and N. Nagaosa,
                   \emph{Science} {\bf 288}, 462 (2000). 

\bibitem{Fei99} L. F. Feiner and A. M. Ole\'s,
                     \emph{Phys. Rev. B} {\bf59}, 3295 (1999).

\bibitem{Dag04} M. Daghofer, A. M. Ole\'s, and W. von der Linden, 
                   \emph{Phys. Rev. B} {\bf 70}, 184430 (2004).

\bibitem{Kha01} G. Khaliullin, P. Horsch, and A. M. Ole\'s, 
                   \emph{Phys. Rev. Lett.} {\bf 86}, 3879 (2001).

\bibitem{Zen51} C. Zener, \emph{Phys. Rev.} {\bf 82}, 403 (1951).

\bibitem{Bri99} J. van den Brink and D. I. Khomskii, 
                   \emph{Phys. Rev. Lett.} {\bf 82}, 1016 (1999).

\bibitem{Woh06} K. Wohlfeld and A. M. Ole\'s, 
                   \emph{Phys. Stat. Sol. (b)} {\bf 243}, 142 (2006).

\bibitem{Kot86} G. Kotliar and A. E. Ruckenstein, 
                   \emph{Phys. Rev. Lett.} {\bf 57}, 1362 (1986).

\bibitem{Fei05} L. F. Feiner and A. M. Ole\'s, 
                   \emph{Phys. Rev. B} {\bf 71}, 144422 (2005). 

\bibitem{Miz96} T. Mizokawa and A. Fujimori, 
                   \emph{Phys. Rev. B} {\bf 54}, 5368 (1996).  

\bibitem{Dag01} E. Dagotto, T. Hotta, and A. Moreo,
                  \emph{Phys. Rep.} {\bf 344}, 1 (2001).

\bibitem{Fuj05} J. Fujioka, S. Miyasaka, and Y. Tokura,
                        \emph{Phys. Rev. B} {\bf 72}, 024460 (2005).

\end{thebibliography}
\end{document}